Leveraging Software, Advocating Ideology: Linux, Free Software, and Open Source

TPRC Draft version
10-21-01

Matt Ratto
Doctoral Candidate
Department of Communication
University of California, San Diego
La Jolla, CA 92093-0503
mratto@ucsd.edu

**Preface**

The Supreme Court has ruled that while "it is possible to find some kernel of expression in almost every activity a person undertakes…such a kernel is not sufficient to bring the activity within the protection of the First Amendment."[1] Dan Burk states "embedded technical expression is at best the sort of kernel described by the Supreme Court, and not the type of expression shielded by the First Amendment."[2] Given the complex nature of regulating technologies, Burk's statement – and the classic liberal separation between expression and action or 'word' and 'deed' it reveals- seems to be a necessary corollary to First Amendment law. However, maintaining a separation between the 'action' or functional aspects of software, i.e. its 'technical expression' , and the 'words' about software, more traditionally expressive aspects may be more difficult for other governance issues given the way functionality is used in the construction of software as socially meaningful.

Whether or not the kinds of expression manifested by software are worthy of First Amendment protection, I leave to First Amendment scholars. Instead, I explore how it is that technologies are used to express opinions about the world. More specifically, I detail how a software program called Linux becomes expressive and meaningful through its fucntionality as well as the way this functionality is leveraged in traditionally discursive practices such as writing and speeches[3].

I argue in this paper that in order to understand how software is socially meaningful, we must understand how software functionality gets leveraged to articulate and defend ideas. This also means reexamining the classic liberal distinction between word and deed. I do not want to argue[4] that software does away with this separation. Regulating a software program such as Napster or DeCSS may often mean characterizing different aspects of it as expressive or as functional. However, understanding such programs requires putting such distinctions aside. In other words, while we may find characterizing certain aspects of a software program as functional behavior or as expressive behavior, (or, say, separating source code and object code,) a necessary step towards regulating them, we must also understand that such distinctions are always tentative.[5]

In section I, I review some recent legal scholarship about software and explain why understanding how embedded technical expression works is important given the increasingly large role of software in governance. In section II, I explore software construction as a combination of social and technical practices. In section III, I describe some examples of meaning-making around the software program Linux. In section IV I conclude with a short description of why such analyses are important.

---

[1] City of Dallas v. Stanglin, 490 U.S. 19, 25 (1989)
[2] Dan Burk,'Patenting Speech', Texas Law Review, Nov. 2000
[3] Two important points here: First, while in this paper I emphasize the programming decisions that go into the construction of Linux as socially meaningful, I also see the adoption of specific licenses and other formal contracts(e.g. the GPL) as equally constructive. Second, the argumentative strategy I use in the paper is to demonstrate the process by which functional aspects of software become expressive through their use in more traditionally discursive forms. The danger here is to see the engineering of function as always primary and its use in discourse as secondary. I would argue that while this order is sometimes the case, the act of constructing software as socially meaningful involves a co-constructive process by which conversation and engineering happen simultaneously. For more on this, see the section below titled "Construction as social, technical, and iterative".
[4] For the dangers inherent in this position, see Lawrence Lessig, "Foreword.", *Stanford Law Review* v52, n5 (May, 2000):987
[5] I would also add that the act of classifying something as functional or expressive is often a strategic move.

**Section I:**

## *Recent legal scholarship*

The study of computer technology within legal scholarship has typically concentrated on intellectual property. Debates about the effects of the computer revolution on IP can be traced back to a series of cases and conferences beginning in the 1960's between law-makers and a developing computer industry. Often this debate centered around how to characterize computer software - as text or as machine - in order to 'fit' software into existing IP regimes of copyright or patent.

More recently, some legal scholars have focused on the wider ramifications of computer technology, particularly in reference to governance. While an overview of this work is beyond the capabilities of this paper, two insights stand out. These are:

- Software has the capability to supercede legal mechanisms of governance, particularly in the arenas of privacy and intellectual property policy. In one sense, computer technology can become a 'lex informatica' that governs in place of more traditional forms of law.[6]

- Since the development of computer software currently happens primarily within the private sphere of for-profit corporations, 'lex informatica' thus creates the possibility of a move in the control of governance from public to private hands. Lessig has characterized this as a move from 'East Coast law' to 'West Coast code'.[7]

These insights demonstrate the importance of understanding the impact of computer technology on governance. Unfortunately, legal scholarship about computer technology has often been hampered by its past attempts to define computer software as text or machine. These attempts have typically focused on considering first whether software is expressive or functional, and second, what aspects of software can be considered to be one or the other. Such distinctions, while necessary in order to 'fit' software into existing IP regimes, make understanding software as governance particularly difficult. Tracing the connection between software function and expression is essential if we want to understand how 'code' works as law.

To assume that I might work out how 'code' works as law within this paper is unthinkable hubris. Instead, I want to use the rest of this paper to articulate a few related points. The first is that legal scholars who study technology, and Science, Technology, and Society (STS) scholars might find cross-disciplinary interaction beneficial. While STS as a field is relatively new, scholars who study STS issues have historically addressed questions similar to the ones facing legal scholars now. In addition, the questions posed by legal scholars about technology and society can provide a necessary grounding for the often more nebulous and philosophic work generated by STS.

One such insight from STS is the social nature of technology. While technical decisions are often defended as rational, logical, or pragmatic, the construction of technological apparatus is the result of a complex interwoven process that includes both social and technical factors. In this case, 'construction' refers not just to the actual coding decisions and program-writing by which software programs are made, but also to the rhetorical and argumentative strategies that 'construct' some software programs as 'better' or as 'more efficient'.

---

[6] Joel Reidenberg, "Lex Informatica: The Formulation of Information Policy Rules Through Technology" in *Texas Law Review*, Feb. 1998.

[7] Lessig, Lawrence. *Code and other laws of cyberspace*, Basic Books, c1999

**Section II.**

*Construction as social, technical, and iterative*

One way to characterize technologies is as socio-technical networks. A number of insights, often collectively called Actor-Network Theory (ANT), is a result of the exploration of technology by such scholars as Michel Callon, Bruno Latour, John Law and others. While many of its original proponents have since rejected some of the precepts of ANT, the issues addressed by this debate serve as a useful method for examining the strategic nature of technological decision-making. Without a critical understanding that sees software construction as both social and technical, software has a tendency to become a 'black box' in which the social decisions disappear and are replaced by seemingly natural, rational, and 'scientific' determinations.[8]

Software programs are powerful argumentative tools because they combine social and technical practices. In addition to the engineering practices of bug fixes, new versions,[9] and the programming of new features and capabilities - i.e. the iterative nature of software development, new arguments are made about the technology, new meanings are ascribed and explored, new social relations are manifested. Importantly, these two parts, the social and the technical, are combined – one is used to defend the decisions made in the other. In part because of this blending, traditional forms of social analysis such as semiotics or rhetoric tends to have limited applicability. Thus, in order to understand how software becomes meaningful in more than just an operational sense, we must examine how its technical characteristics get used to articulate social decisions as well as how social decisions get embodied within the technical apparatus.

Therefore, it order to understand how software 'works' in and on the world, we have to understand both its functionality, i.e. its technical expression, as well as examine how this functionality gets used as argument.

In this paper I focus on the software program Linux. I do this for two reasons. First, Linux serves as a good example of a program that has come to stand in for larger social debates about the Internet, about digital information, about a 'New Economy', and about associated distributed work practices. In addition, much of the debate about Linux – as with some other software programs – has increasingly taken on a kind of moral tone in which decisions made about the use of Linux or the adoption of the organizational processes that accompany it are often defended as morally good or bad.Such discursive strategies make analyzing the blending of the social and the techncial much easier to examine. While I do see all programs as including social and techncial constructive practices, examining those of, say, Microsoft Excel which appears[10] to be less socially loaded, would be more difficult.

---

[8] for more on technology, socio-technical networks, and Actor-Network Theory see *Actor network theory and after /*, edited by John Law and John Hassard. Oxford [England] Malden, MA : Blackwell/Sociological Review, 1999.

[9] I would also include the coding of program cracks (code created to 'crack' the software's embedded use restrictions), and program 'hacks' (code created to extend the program) despite being carried out without the permission of the orginal programmers. In fact, some program developers have embraced the hacking of their programs ( albeit in limited ways) by creating tools that allow users to create new 'skins' (interface elements) or new character behavior (in the case of video games.) Examples include such programs such as WinAmp (a MP3 player) and Quake (a popular video game.)

Second, Linux serves as a good case to explore the issue of what makes embedded technical expression meaningful because of the public nature of its construction process. Since Linux is an 'open software' project, it becomes possible to view listserve posts between developers debating engineering decisions, public 'manifesto' documents that attempt to define ( or redefine) the norms of the developer community,  as well as the formal licenses that mediate developer and developer-user relationships. Proprietary software, (such as Excel) have much more closed development processes making access (and analysis) more difficult. Thus, Open Code programs provide good examples because the social and the technical processes of their construction are more available for analysis – this is one good reason why such development forms may be a more accountable, if not more democratic approach to software development.

---

[10] I say 'appears' because it may turn out that seemingly pragmatic and rational programs like Excel, upon analysis, turn out to be just as socially loaded as Linux. For why such an analysis would be difficult, see below.

**Section III.**

## *Linux as OS and argument*

### Linux as operating system

So what is Linux? At one level Linux is a software program that can facilitate the operation of computer hardware. This type of program, called an operating system, provides a layer of abstraction between the computer hardware and the application software that the computer user actually uses to accomplish tasks. Without an OS, each piece of application software would have to directly address the internal workings of the computer, an almost hurculean task given the mutiplicity of differences inherent in even the most similar of computer systems. More specifically, Linux is the kernel of an operating system made up of a number of different software programs that together accomplish the functions of an OS. While the kernel of an OS is only part of the overall code necessary to support a computer, it is, by far, the most complex and necessary part. While an OS might function with many of its pieces missing, it cannot function without a kernel.[11]

While the label 'Linux' more appropriately refers to the kernel software that resulted from the development effort started by Linus Torvalds, it has come to refer to the many different collections of software that together with the Linux kernel come to form a complete OS. These collections, assembled and distributed by a number of different profit and non-profit organizations are collectively called 'Linux distributions.'

In another sense, Linux is an artifact that often stands in as the result of (and an argument for) a larger argument about organization, work, property, and a new, distributed economy. This argument, however, does not consist of a singular perspective. While there are many different voices within it, two different but related terms, Free Software and Open Source software, constitute the major viewpoints. Before addressing this debate, let me provide some background into the development of Linux.

### Linux background

One mostly agreed upon point of origin is Linux Torvald's often quoted post to the minix listserve in October of 1991.

> Do you pine for the nice days of minix-1.1, when men were
> men and wrote their own device drivers? Are you without a
> nice project and just dying to cut your teeth on a OS you can
> try to modify for your needs? Are you finding it frustrating
> when everything works on minix? No more all-nighters to
> get a nifty program working? Then this post might be just for
> you :-)[12]

In this post Torvalds, a student in computer science at the University of Helsinki, asked for contributions to his project, the coding of a operating system based on the design of Unix.

---

[11] See Appendix A for overview
[12] posted to info-mini@udel.edu; from:Linus Benedict Torvalds; subject: Free minix-like kernel sources for 386-AT

Torvalds started what was later to be called Linux in part because of the desire to experiment with his new Intel 386 computer and to learn 'protected mode' and other programming techniques possible by the 386 architecture. Also, like many other users, Torvalds was dissatisfied with the Minix operating system he was currently using. Minix was, like Linux was later to be, a partial clone of Unix made for Intel-based personal computers. It had been developed by Andrew Tannenbaum at the Free University of Amsterdam in order to teach operating system programming.[13] However, Minix was more a teaching tool than a full-fledged operating system. For years, Tannenbaum had resisted efforts to extend Minix because he was trying to keep its code base small enough to be covered in a one semester computer course. Torvalds decided to leverage his knowledge of Minix and the knowledge of the preexisting Minix community to create a 'new' Unix clone.[14] Fortuitously, a previous Unix clone effort had already written many parts to a full featured Unix system. This effort, started by Richard Stallman, was called the GNU project for 'Gnu's Not Unix'. The GNU project was, in a way, very successful, although Stallman was never able to fully satisfy his goal of a 'complete' free Unix system. But the code that GNU community was to create over the next decade was to prove invaluable to the Linux effort. More importantly, Stallman implemented a licensing scheme he called 'copyleft' within the General Public License (GPL) used to license GNU software. 'Copyleft' is a kind of copyright agreement that allows users to sell, copy, and change GNU software - but the same rights have to be passed along to the following users. Further, copyleft requires that the source code necessary to modify software must be included in all distributions.

Torvalds decided to code a Unix clone kernel that could use the existing GNU code base. By combining Linux(the kernel) with the GNU C compiler, shell programs, and editors, a complete operating system could be created. Further, by eventually licensing Linux as 'copyleft' software, Torvalds encouraged Linux to be collaboratively extended and developed.

Linux can thus be seen as both an individual and a collaborative project which used exisiting labor resources (the Minix community among others) as well as an exisiting code base (the GNU code and license) to create a free software program capable of competing with the commercial software industry's best efforts. Currently (as of June, 2001) Linux is supposedly accounts for around 30% of the servers which operate on the Internet[15]. In addition, many commercial computer companies either distribute and sell Linux distributions (RedHat, Mandrake, Corel, etc…), use pre-installed Linux computers as part or all of their product line, (Penguin Computing, Hewlett-Packard, IBM, etc…), or develop software solutions that make use of the Linux operating system. Development continues on the Linux kernel (still headed by Torvalds) as well as on many other system components. Such efforts are sometimes more or less open public projects, and sometimes privately funded projects. Thus Linux is the result of a large number of distributed efforts, both public and private, that have culminated in a series of system components, the most important of which is the Linux kernel proper, that can be combined in a number of different ways to create multiple distributions of Linux that are given away for free or commercially sold.

---

[13] Tannenbaum began developing Minix in 1987 after AT&T, then the owners of Unix, increased its licensing costs making it too expensive to use as a teaching tool.

[14] For Tannenbaum's reaction to Linux and the debate that followed see "The Tanenbaum-Torvalds Debate" in *Open Sources: Voices from the Open Source Revolution* / Edited by Chris DiBona, Sam Ockman & Mark Stone. O'Reilly Press, 1999

[15] According to http://www.netcraft.com/survey/

*Linux as argument*

Currently, one of the most important debates about Linux involves its use to defend the theoretical assumptions of two connected but contradictory development positions - the debate between 'free software' and 'open source software'. The most public faces of these positions belong to two important figures of public software development, Richard Stallman and Eric Raymond. This debate in part revolves around a conception of Linux. More importantly, this debate reveals the use of the functionality of Linux to defend ideological positions.

Stallman and the FSF

Richard Stallman is the founder of the Free Software Foundation, an organization created to facilitate the development of a software programming effort called 'GNU', for 'Gnu's not Unix. Unix, then owned by AT&T, had recently dramatically increased the costs of licensing and using Unix. In response to this (and what he felt was the increasing commercialization of software more generally), Stallman decided to create GNU, a free version of Unix.

By 1991 the FSF, led by Stallman, had created almost every part of a Unix-like operating system. More importantly, Stallman implemented a licensing scheme he called 'copyleft' within the General Public License (GPL) used to license GNU software. 'Copyleft' is a kind of copyright agreement that allows users to sell, copy, and change GNU software - but the same rights have to be passed along to the following users. Further, copyleft requires that the source code necessary to modify software must be included in all distributions.

As noted above, in 1991 Torvalds decided to license Linux under the GPL. The relationship between Stallman and the Linux development community has been stormy at best. This is partly due to Stallman's attempt to rename Linux as Linux/GNU (or more abrasively as LiGNUx) because of the way the Linux operating system makes use of GNU-created software, most importantly the GCC compiler – an essential tool. However, Stallman has not hesitated to point to Linux as a primary example of the success of free software development methods.

Raymond and the OSI

Eric Raymond is the author of a number of different papers on software development and distributed forms of programming work. The first of these papers, 'The Cathedral and the Bazaar', was first given as a talk at a Linux conference in 1997. Raymond's article, since available on the web (and more recently published as a book by the same name by O'Reilly Press) had a dramatic effect on the software industry.[16].

---

[16] One example is Netscape Corporation's decision in 1998 to release the source code for their web browser software see Netscape's press release at http://www.netscape.com/newsref/pr/newsrelease558.html as of 21-Aug-2000.

In 'The Cathedral and the Bazaar' Raymond sets out two software development models, the 'cathedral' and the 'bazaar'. After setting out his credentials as an open source developer[17], Raymond admits that up until Linux, he had thought that:

> ...important software (like operating systems and really large tools...) needed to be built like cathedrals, carefully crafted by individual wizards or small bands of mages working in splendid isolation.... [18]

The success of Linux caused him to rethink this theory.

> Linus Torvald's style of development -release early and often, delegate everything you can, be open to the point of promiscuity - came as a suprise. No quiet, reverent cathedral building here - rather, the Linux community seems to resemble a great babbling bazaar of differing agendas and approaches (aptly symbolized by the Linux archive sites, who'd take submissions from anyone) out of which a coherent and stable system could seemingly emerge only by a succession of miracles.[19]

In 1998, Raymond and a few others created the 'Open Source Initiative', an association aimed at interesting the high tech industry in open source development methods. As the history page on its web site says:

> We were reacting to the Netscape's announcement that it planned to give away the source of its browser. One of us (Raymond) had been invited out by Netscape to help them plan the release and follow on actions. We realized that the Netscape announcement had created a precious window of time within which we might finally get the corporate world to

---

[17] Raymond has been involved with 'free' software development for many years, including the development of a 'free' mail program called 'fetchmail'. He uses the 'fetchmail' development case to analyze and defend his notions of 'open source' development in 'The Cathedral and the Bazaar'.

[17] Raymond, Eric 'The Cathedral and the Bazaar' at http://www.tuxedo.org as of 21-Aug-2000.

[17] Ibid.

[17] at http://www.opensource.org/history as of 21-Aug-2000

[17] Salon Magazine, 'Let my Software Go!' at http://www.salon.com/21st/feature/1998/04/cov_14feature2.html as of 21-Aug-2000.

[17] from 'Why 'Free Software' is better than 'Open Source' at http://www.gnu.org/philosophy/free-software-for-freedom.html as of 21-Aug-2000.[17] More about X windows and the consortium and license here.

17 Servers are computers which provide basic services to other computers on a network. Linux-based computers are often used to 'serve' web pages, as printer servers, or as other lower level components of a computer network. Linux computers in these roles are mostly maintained by engineers or programmers and thus have not had to face many of the same problems 'desktop' computers face.

17Among them are AT&T Unix and other Unix clones such as FreeBSD and NetBSD.

17 I should also mention that not all programmers interested in making Linux 'more approachable' are in support of the commercialization of Linux.

[17] While software code, behavior, and architecture exist as a kind of materialization of relationships, analyzing these relationships can easily become a futile exercise if done apart from the software's actual use by involved communities of programmers or users. Such situated analy
[e]s, although necessary, are often difficult and time-consuming. For the purposes of this pa
[e]r I wil

> listen to what we have to teach about the superiority of an
> open source development process.[20]

This last statement demonstrates the source of conflict between 'Open Source' ( Raymond) and 'Free Software' (Stallman). For Raymond, open source development represents a development form that results in faster development times and better software. For Stallman, however, 'free software' is not just about better software, but is also about freedom and ownership. This difference can be summarized by the most quoted aphorisms of both. Raymond's aphorism refers to development and software testing: 'Given enough eyeballs, all bugs are shallow.' Stallman's aphorism refers to property and ownership 'Free software is like free speech, not free beer.'

In an interview with Salon magazine, Raymond lay out the differences between himself and Stallman:

> I love Richard dearly, and we've been friends since the 70's
> and he's done valuable service to our community, but in the
> battle we're fighting now, ideology is just a handicap. We
> need to be making arguments based on economics and
> developmental processes and expected return. We do not
> need to behave like Communards pumping our fists on the
> barricades...[21]

Stallman explains the difference as:

> The main argument for the term 'open source software' is that
> 'free software' makes some people uneasy. That's true:
> talking about freedom, about ethical issues, about
> responsibilities as well as convenience, is asking people to
> think about things they might rather ignore. This can trigger
> discomfort, and some people may reject the idea for that. It
> does not follow that society would be better off if we stop
> talking about these things.[22]

What is interesting about the debate between Stallman and Raymond is that it is not specifically about the functionality of Linux, i.e. what tasks the software should include, but is instead about what kinds of social relations Linux demonstrates. However, it should be obvious that the statements of Stallman and Raymond could not be about just any software. While their perspectives are not about Linux functionality, they rely on it in order to foster and defend their positions. Therefore, in order to understand how social relations are constituted around Linux, we have to begin to examine how Linux itself expresses these relationships.

### *Code, Behavior, Structure*

---

[1] concentrate on the use of code, behavior, and architect

ones such as FreeBSD and NetBSD. Current numbers are:?

[22] I should also mention that not all programmers interested in making Linux 'more approachable' are in support of the commercialization of Linux.

To return to my main argument, this paper uses the software program Linux and the discourse around it in order to examine how software programs can be used to articulate and defend social and economic positions. Although I do not use the term 'expression' in a strict legal sense, I claim that in order to make policy decisions involving software, it is important to understand how the functionality of software is expressive. Another way to state this is that software programs like Linux are socially meaningful through functionality and talk about functionality.

In the section above, I briefly summarized the perspectives of the main advocates of two related but different perspectives on Linux. Both positions attempt to define what is meant by Linux by leveraging the same functionality. But understanding their arguments mean breaking down what we mean by functionality. More importantly, it means understanding how functional characteristics can be leveraged to make social arguments.

An initial step is to overcome the problematic dichotomy of interface/engineering that sits at the heart of a common-sense understanding of software. Although this separation makes sense on the surface, it often results in social analyses of software programs that focus on the behavior of the program, i.e. interface elements, while ignoring the code itself. Worse, maintaining the interface/engineering dichotomy can result in an analysis that views the interface aspects of the program as social and expressive, while seeing the engineering aspects as purely functional and rational. A possible way to overcome this problem is to characterize software differently. Below I lay out a possible solution and provide some examples based on Linux.

Software is technically expressive in three ways;

- As source code that expresses ways to accomplish specific programming tasks, i.e. relationships between programming and programmer;

- Through the behavior of the program that constrains the actions of its users, i.e. relationships between tasks and users;

- Through the structure of the code that organizes the modes of labor by which the program can be built and/or extended by multiple people, i.e. relationships between programmers.

Expression through source code:

This aspect of the expressive nature of software appears to fit most securely into the traditionally legal notion of expression. In the few software cases that have attempted to defend the distribution or creation of software programs under the rubric of First Amendment rights (Berstein, Junger, Karn) the proceedings have focused on how the source code of the program in question has expressed methods of programming. In these cases, the defense of the programs in question as expression have typically involved citing the commentary of programmers embedded in the source code as well as the way the source code expresses ways of accomplishing programming tasks. The courts' decisions have typically been based on whether or not the program is primarily expressive or primarily functional, the purpose of the expression cited (i.e. whether or not it was political speech), and whether or not the expressive quality of the source code would be understood by a possible audience.

While the court decisions to date remain divided whether or not source code meets the expression requirements necessary to be afforded First Amendment protection, it seems quite obvious that source code is expressive. As the Berstein court stated, "The distinguishing feature of surce code is that it is meant to be read and understood by humans and that it can be used to express an idea or a method'.[23]

There are many examples of programmers debating programming methods and ideas within the Linux development community. Below is one example from the linux-kernel development listserve. As in many email discussions, the current poster (or the software he is using) marks off the writing of others by prefacing it with a '>'.

```
XXXX-XXXXX writes:
> if the smp_function area is busy, then smp_call_function() executes the
> following code:
>
> > if (retry) {
> > while (1) {
> > if (smp_call_function_data) {
> > schedule (); /* Give a mate a go */
>   ^^^^^^^^^
> > continue;
> > }
> > spin_lock (&lock);
> > if (smp_call_function_data) {
> > spin_unlock (&lock); /* Bad luck */
> > continue;
> > }
> > /* Mine, all mine! */
> > break;
> > }
> > }
>
> Is that ok?

I think it's fine (I'm the author, BTW).
```

In this email conversation, two programmers on the Linux-kernel listserve are using examples of source code to illustrate and debate programming methods. Of particular interest is the tongue-in-cheek nature of the comments (within the '/*' and '*/') and the way they rely on an understanding of the code in order to translate their meaning. This code snippet is run when another part of the program (the smp_function) is in use by another section of the OS. This can occur because the Linux OS has the capablity of running multiple tasks simultaneously (multi-tasking). Because of this, the program must have the capability to negotiate these separate processes (called threads). In the example above, the smp_call_function executes when the smp_function is asked to run but is already in use. It sets up a loop which checks if the smp_function is free ('Give a mate a go'). If it is not free, it continues waiting('Bad luck') until it becomes free and then continues on ('Mine, all mine!'). Understanding the code requires simultaneously following the code and connecting the comments, (though in this case the code is relatively simple and the comments not necessarily illustrative.)

---

[23] Bernstein, 176 F.3d at 1140

The post goes on:

```
> e.g what if the current thread is a
realtime thread, could this
> deadlock? Is is possible that
schedule() will never return?

The current thread *is* the one calling
smp_call_function().
Or are you thinking about calling
smp_call_function() from outside a
process context? You're not supposed to
do that. Therefore it's not a
problem ;-)
```

Here the first poster describes a scenario that might cause the code to break. The current author denies that the code would 'deadlock' by pointing to the inappropriateness of the original author's scenario, i.e. that the code snippet would be called 'from outside the process context'. The code is thus 'expressive' of coding practice in two related ways; first by 'breaking' if used incorrectly, and second through its use in discourse between programmers to define correct programming activity.

Expression through program behavior:

This aspect of the expressive nature of software is much more problematic in regards to the legal definition of expression. In fact, the type of expression manifested by the behavior of computer programs is exactly the kind of expressive 'kernel' that is most probably not protectable as expression.[24] However, it is important to recognize the way software program behavior communicates relationships between users and tasks by providing or constraining uses, just as the source code organizes relationships between programmers and acts of programming described above.

All tools 'afford' certain uses by the way the reward or punish the actions we take with them.[25] For example, a hammer provides a way of driving nails that rewards certain ways of accomplishing this task, (e.g. holding the handle and driving the nail using the head of the hammer rather than the opposite) software programs have affordances that effect the kinds of practices users do. Computer programs do the same. Microsoft Word, for example, structures the act of writing a letter in part by automating the process. While users can turn off this functionality, the embedding of a specific letter-writing process in the software explicitly expresses the embedded process as authoritative. Further, Word then rewards users that use the automatic letter-writing process by shortening the time it takes to format a letter, albeit in the form the programmers of the Word software have decided is appropriate. In this way, software programs express appropriate methods of doing tasks to users, while not necessarily constraining user behavior.

---

[24]See above
[25]Gibson, J. J. "The theory of affordances". In R. E. Shaw & J. Bransford (Eds.), *Perceiving, Acting, and Knowing*. Hillsdale, NJ: Lawrence Erlbaum Associates.(1977)

Linux is no different from other tools in this regard. In fact, one of the ongoing debates about Linux involves the way it affords certain types of activities by specific groups of users.

Previously thought of as a 'geeks only' operating system, Linux has started to change. While Linux as server software[26] has long been challenging Windows NT and other operating systems[27], only recently has it begun to move towards the desktop, attempting to replace such operating systems as Mac OS and Windows 2000. The term 'moving Linux to the desktop' refers to the attempts by Linux advocates to encourage the use of Linux as an operating system for personal use.

While there are still some Linux advocates who actively resist what they see as the ongoing commercialization of Linux, making Linux 'more approachable' is seen as a worthy aim by most Linux programmers[28]. An example of this trend is the GNOME project, originally formulated by Miguel de Icaza at the University of Mexico, and since adopted as the primary graphic desktop for Linux by many software development companies such as Redhat and others. GNOME provides a graphic interface for Linux which can, in addition to its own interface, also mimic the look and feel of both Macintosh or Windows computer software.

Graphic interfaces for Linux that attempt to facilitate its adoption by providing familiar interfaces to new users is just one example of how program behavior organizes relationships between users and software. Other examples include programs that simplify the complex installation or updating process of Linux. These tools, such as Redhat's RPM package manager or Mandrake's graphic installation program, attempt to include more users by providing ways to make Linux adoption easier.

## Expression through code structure

The concept of structure as expressive of social values has recently been addressed by a number of scholars within the social study of science and technology. Such work has examined the structure of computer programs used in hospitals(Berg), the structure of systems used to classify medical disorders (Star and Bowker), and the structure of large-scale biodiversity databases. (Bowker). In the context of the discussion above, structure refers to the organization of different code elements within the overall software program. For example, the concept of Object Oriented Programming (OOP) refers to a programming structure in which individual software processes are organized into discrete software 'objects'. Such objects can be used multiple times within the same software program, and also makes the 're-use' of code in other programs much simpler. In addition, most large software programs are organized into such categories as executables, libraries, and databases. Such organizing code structures also organize the labor that goes into coding, maintaining, or extending programs. In this way, code structure expresses relationships between programmers.

---

nts of a computer network. Linux computers in these roles are mostly maintained by engineers or programmers and thus have not had to face many of the same problems 'desktop' computers face.
[27] Among them are AT&T Unix and other Unix clones such as FreeBSD and NetBSD.
[28] I should also mention that not all programmers interested in making Linux 'more approachable' are in support of the commercialization of Linux.

An important structural feature of Linux derives from its use of Unix as a structural model. In the Unix model, the operating system is made up of a series of different programs responsible for different functions, coordinated by a core software program called the 'kernel'. As noted above[29], the Linux OS follows this model. One example is the separation of Linux graphic windowing code from the core OS. The most prevalent software code that creates the graphic interface for Linux is called X-windows and is developed and maintained by a separate group of programmers.[30] This division allows different groups of programmers to work together on the overall OS without an overall coordinating authority. Different programmers can create different parts without having to worry too much about interoperability – as long as they can agree upon the standards imposed (in part) by the OS kernel.

Within the Linux development community, the individual programming efforts that create the different aspects of the overall OS are mostly left to their own devices. As long as the resultant code can 'talk' to the kernel and does not break any of the standards upon which Linux is based, they are adopted into the different Linux distributions assembled by the different profit and non-profit Linux distributors.

Thus the structure of the Linux distributions, collected around the Linux kernel, but including different OS 'pieces' collected by Linux distributors, organizes relationships between programmers somewhat differently than most other OS's. The disassociated nature of the overall OS means that coordination between programmers can be loosely organized.

Such freedom is less true of the Linux kernel development effort in which inclusions of new code within the kernel is controlled by a collection of gate-keepers with Torvalds himself as the ultimate authority. Still, even the kernel code structure itself organizes certain types of relationships. For example, in 1995, Linus Torvalds with the support of a core group of kernel programmers, rewrote the current Linux code in order to make it able to load different kernel 'modules'[31] depending on what functionality was required. While this was in part due to a desire to make Linux more portable (meaning more able to be 'ported' to run on different computer hardware) this change in the structure of the Linux kernel also facilitated a more diverse kernel development community. With the addition of separate kernel modules, Linux developers could create new kernel-specific code (such as ethernet drivers, file systems, or hardware-specific code that must be run as part of the kernel) without having to go through Linus Torvalds.

Another important aspect of dynamic modules points even more explicitly to the way code structure expresses relationships. Torvalds' decision to license the Linux kernel under the GPL (GNU Public License) caused problems for software developers who wanted to create custom code for Linux, but did not want to made their software publically available. Under the GPL, any works 'derived from' GPL'd software were themselves automatically covered under the GPL. Stallman wrote the GPL explicitly to do this in order to prevent proprietary software developers from taking and using the open source code of GPL'd works without giving the resultant 'derived' code back to the GPL community. But this restriction created a problem for hardware manufacturers who wanted to create products for Linux computers (or wanted to make their existing products work with Linux), but didn't want to make the necessary driver software source code public. While still a grey area within the GPL,

---

[29] See section titled 'Linux as OS'.

[31] For more information, see Matt Welsh, "Implementing Loadable Kernel Modules for Linux", *Dr. Dobb's Journal*, May, 1995. on the web as of 10/01 at http://www.ddj.com/articles/1995/9505/9505a/9505a.htm

Torvalds decided that dynamically linked modules would not be considered 'derived works' as defined by the GPL. Thus, proprietary developers could create driver modules for Linux without having to make their module code public.[32]

As detailed above, the structure of Linux code makes possible the distributed character of the Linux development process, as well as the inclusion of both profit and non-profit, proprietary and open source code. In this sense, Linux code structure organizes relationships between programmers as well as expresses these relationships as possible and even normative.

### *Functionality as doubly expressive*

Source code, program behavior, and code architecture express different relationships via their functionality. In addition, as the Stallman and Raymond example demonstrates, these elements are used in more traditionally discursive forms such as speeches and writings to express value judgements about these relationships. Therefore, software is in a sense doubly expressive; first by authorizing specific relationships between programmers, users, the code, and tasks; second by providing examples that can be used to articulate, defend, or characterize forms of these relationships[33]. Another way to state this is that software as code, behavior, and structure both *expresses* relationships as well as *organizes* those relationships.

---

[32] For Torvalds' own 'take' on this, see Linus Torvalds, "The Linux Edge', in *Open Sources: Voices from the Open Source Revolution*, DiBonna ed., O'Reilly Press.

[33] While software code, behavior, and architecture exist as a kind of materialization of relationships, analyzing these relationships can easily become a futile exercise if done apart from the software's actual use by involved communities of programmers or users.

**Section IV.**

## *What does this mean for policy makers?*

Having laid out a possible framework for understanding how software functionality expresses and organizes relationships, I now want to point to why analyzing functionality is an important step to understanding the strategic use of software. Microsoft's recent strategy in regards to Linux is a good example.

### Microsoft, 'viral software', and 'de-commoditizing the protocols'

Microsoft has recently begun an increasingly vocal campaign against Linux and open source/free software more generally. One Microsoft executive, Jim Allchin, was quoted as saying that "Open source is an intellectual-property destroyer. I can't imagine something that could be worse than this for the software business and the intellectual-property business."[34] Another, Steve Ballmer, in an interview with the Chicago Sun-Times said, "Linux is a cancer that attaches itself in an intellectual property sense to everything it touches."[35]

More recently, Microsoft released a new license for a beta version of their Mobile Internet Toolkit, a software program used by developers to write server software to connect with handheld computers over the Internet. This license prohibits developers from using the software "...in conjunction with Potentially Viral Software," which Microsoft defines as "... any software that is distributed as free software, open source software (e.g. Linux) or similar licensing and distribution models..."[36]

Microsoft's strategy to delegitimate Linux by defining it as 'viral software' was not received well by open software advocates, one who, responding in kind, said:

The GPL is not a virus, it is a vaccine, an inoculation against later abuse of your code by having someone, such as Microsoft, take your hard work, incorporate it into a proprietary product which is then extended and kept closed, marginalizing your project in the process."[37]

This comment reveals the author's familiarity with another Microsoft-related document, the so-called 'Halloween Documents'. These documents are internal Microsoft memos, surreptitiously forwarded to open source advocate Eric Raymond on October 31, 1999, who then made them available on the web.

In the first memo, Microsoft engineer Vinod Valloppillil laid out Microsoft's engineering strategy in regards to Linux. In one section titled "De-commoditize protocols & applications:" he states:

---

[34] http://news.cnet.com/investor/news/newsitem/0-9900-1028-4825719-RHAT.html?tag=ltnc  
[35] http://www.suntimes.com/output/tech/cst-fin-micro01.html  
[36] For an excerpt from the Mobile Internet Toolkit license and a brief analysis of Microsoft's strategy, see Stephen Shankland, "Microsoft license spurns open source", CNET News.com, on the web at http://news.cnet.com/news/0-1003-200-6352301.html.  
[37] Quoted in the CNET article above, originally posted on the open source discussion site, http://www.slashdot.com.

> OSS projects have been able to gain a foothold in many
> server applications because of the wide utility of highly
> commoditized, simple protocols. By extending these
> protocols and developing new protocols, we can deny OSS
> projects entry into the market.[38]

The statement 'de-commoditizing the protocols' requires some unpacking. First, 'commodity' protocols can be understood as similar to market commodities such as wheat or salt.[39] Therefore, commodity protocols are standard protocols that are open, non-proprietary, and available to all. Examples include TCP/IP, DNS, HTML, and most of the protocols that serve as the standards of the Internet. Microsoft's notion of 'de-commoditizing' thus refers to a process of taking existing standard protocols, such as those used in programs like Linux, and slightly exending them in order to make them non-standard and proprietary. This forces developers to choose between developing for Microsoft's platform which uses the now 'de-commoditized' protocol, and other systems that use the commodity standard. By leveraging 'network effects', Microsoft can 'encourage' developers to continue developing for the Microsoft platform, rather than competing systems such as Linux. Therefore, Microsoft's engineering strategy of 'decommoditizing the protocols' can be seen as a more clandestine form of their more public, discursive strategy of naming Linux as 'viral software.' Both serve to extend Microsoft's hold on the software marketplace.

*Conclusion*

The Microsoft example above quite clearly demonstrates the need to understand engineering decisions as often rooted in the same kinds of desires which drive the decisions articulated in more traditionally discursive structures such as speeches, writings, and software licenses.While this is obviously not always true – technical requirements often drive technical decisions – we must not segregate such processes from analysis. Seeing how technologies are constructed to bolster the social requirements of certain interests, to create certain kinds of social realities, and to defend social and economic positions – in other words, to see technologies themselves as part of argumentation – is increasingly important. Understanding technical decision making as not a purely rational or logical process is nothing new. Few engineers would attempt to defend all of their decisions on purely scientific rationale. Equally, the work of many of the legal and information science scholars present at this conference has been to de-rationalize some of the technical decisions made and insubstantiated in digital rights management software, internet protocols, and the like. What I hope the examples I related above do is to rearticulate the need to analyze software programs via the definitions made of them in speeches and writings, but more importantly, by the way functional characteristics of program code, behavior, and structure are leveraged to express and organize relationships between software, programmers, and users. Understanding how software 'governs' is a necessary corollary to understanding how to govern software.

---

[38] Online at www.opensource.org/halloween/
[39] This notion of 'commodity' is quite dissimilar to its use in a Marxist sense to refer to objects disassociated from the material conditions of their production.